\documentclass{article}
\pdfoutput=1
\usepackage{graphicx} 
\usepackage{geometry}
\usepackage{ulem}
\usepackage{authblk}
\def\hhref#1{\href{http://arxiv.org/abs/#1}{arXiv:#1}} 

 \usepackage{hyperref}
\hypersetup{colorlinks,linkcolor=red,urlcolor=blue}

\usepackage{amsmath,amssymb}

\usepackage{multirow}

\title{Pad\'e Approximants for Geodesy}
\author[a]{Ovidiu Costin}
\author[b]{Gerald V. Dunne}
\author[a]{Crichton Ogle}
\affil[a]{Department of Mathematics, The Ohio State University, Columbus OH}
\affil[b]{Department of Physics, University of Connecticut, Storrs CT}
\date{\today}

\begin{document}

\maketitle
\begin{abstract}

    In this note we analyze the use of Pad\'e approximants for downward continuation beyond the radius of convergence of spherical harmonic expansions (SHEs), and for identifying the complex singularities of the gravitational potential. SHEs are, in essence, expansions in 1/r, i.e., expansions about the point at infinity. Their domain of convergence is generically the exterior of the Brillouin sphere. However,  for synthetic models with analytic topography and density the
    region of convergence may be larger, with the deviation decreasing as the structural complexity of the planet  increases.
\end{abstract}

\section{Introduction}
Spherical harmonic expansions (SHEs) of the gravitational potential are expansions in $1/r$, centered at infinity, and with probability one, for celestial bodies, the SHEs converge in the exterior of the Brillouin (Br) sphere and only there \cite{theorem,ogle}. Br is the minimal sphere centered at the barycenter and enclosing the whole planet. For synthetic models with analytic topography and density, Br may be slightly smaller; but this deviation decreases as the structural complexity of the topography and density increases.  

In this paper we analyze the use of Pad\'e approximants to (a) obtain downward continuation of the SHE of the gravitational potential beyond the radius of convergence and (b) to identify the locations of the complex singularities that determine the region of convergence.  The overall motivation arises from the central role played by SHEs in physical geodesy, both analytically and numerically  \cite{M,jekeli,pavlis,hirt,seitz,fukushima}. 

While rational approximation techniques have been employed in numerical geodesy primarily for approximating kernels and integral transforms, their direct application to SHEs for downward continuation appears to be absent from the literature. Pad\'e approximants \cite{baker,bender-book} represent an efficient, easy to implement, analytic continuation mechanism that yields stable continuation below the Brillouin sphere, given sufficient accuracy of the coefficients \cite{pade-noise}.

In recent work \cite{bevis-2024,axi} it has been shown that the convergence properties of the spherical harmonic expansion (SHE) of gravitational functions (e.g. the potential or the gravity field) is governed by the complex singularity structure of the gravitational potential. See also \cite{fukushima}. This fundamental mathematical fact was analyzed in detail in \cite{axi} for the special case of axisymmetric planets, for which there is also a simple geometric method to determine the singularity structure directly from the shape of the planetary topography, without computing actual SHE coefficients. In this paper we provide a complementary perspective on these results, using Pad\'e approximants \cite{baker,bender-book}. 
Recent work combining Pad\'e analysis with conformal and uniformizing maps has been shown to provide analytic continuation with remarkable precision, even with a limited amount of input data concerning truncated Taylor series \cite{plb-pade,cmp-pade,epj-pade}.

In this paper we argue that by combining the singularity perspective with Pad\'e methods, one can obtain high-precision results for gravitational functions well beyond the radius of convergence of the SHE. However, there is also a fundamental limitation to the practical use of Pad\'e methods when the input coefficients of the SHE have limited precision \cite{pade-noise}. These advantages and disadvantages must be balanced.

The main result of \cite{axi} is that for an axisymmetric synthetic planet with constant density, whose cross-sectional boundary profile is given by a parametric curve $(\pm s(z), z)$ in the cross-sectional $(x, z)$ plane, with the $z$ axis being the axis of symmetry, the singularities of the gravitational potential can be deduced directly from the boundary profile function $s(z)$. See figure \ref{fig:axi}, and sections \ref{sec:sing} and \ref{sec:geometric}. 
The radius of convergence of the SHE is determined by the complex singularities of the gravitational potential along the axis of symmetry.
In this paper we show that the same information can be obtained numerically from the coefficients of the SHE:
the Pad\'e poles accumulate to branch points at the location of these complex singularities. Furthermore, Pad\'e is only sensitive to the singularities on the first Riemann sheet. 
This is illustrated here for axisymmetric planets, but the Pad\'e approach can be applied to more general planets, and also provides analytic continuation beyond the radius of convergence.

\subsection{Axial symmetry}

We first review some basic properties of axially symmetric planets, in order to make direct comparisons with the results of \cite{axi}. 
Axial symmetry implies that the gravitational potential $\Phi(r, \lambda, \theta)$, expressed in terms of spherical polar coordinates, is a function of the radial distance $r$ and the colatitude $\theta$, but not the azimuthal (longitudinal) angle $\lambda$. Therefore, the spherical harmonic expansion (SHE) of the gravitational potential can be written as:
\begin{eqnarray}
\Phi(r, \theta)=\sum_{n=0}^\infty \frac{A_n}{r^{n+1}}\, P_n(\cos\theta)
\label{eq:axipot1}
\end{eqnarray}
\begin{figure}[h]
\centerline{\includegraphics[scale=.25]{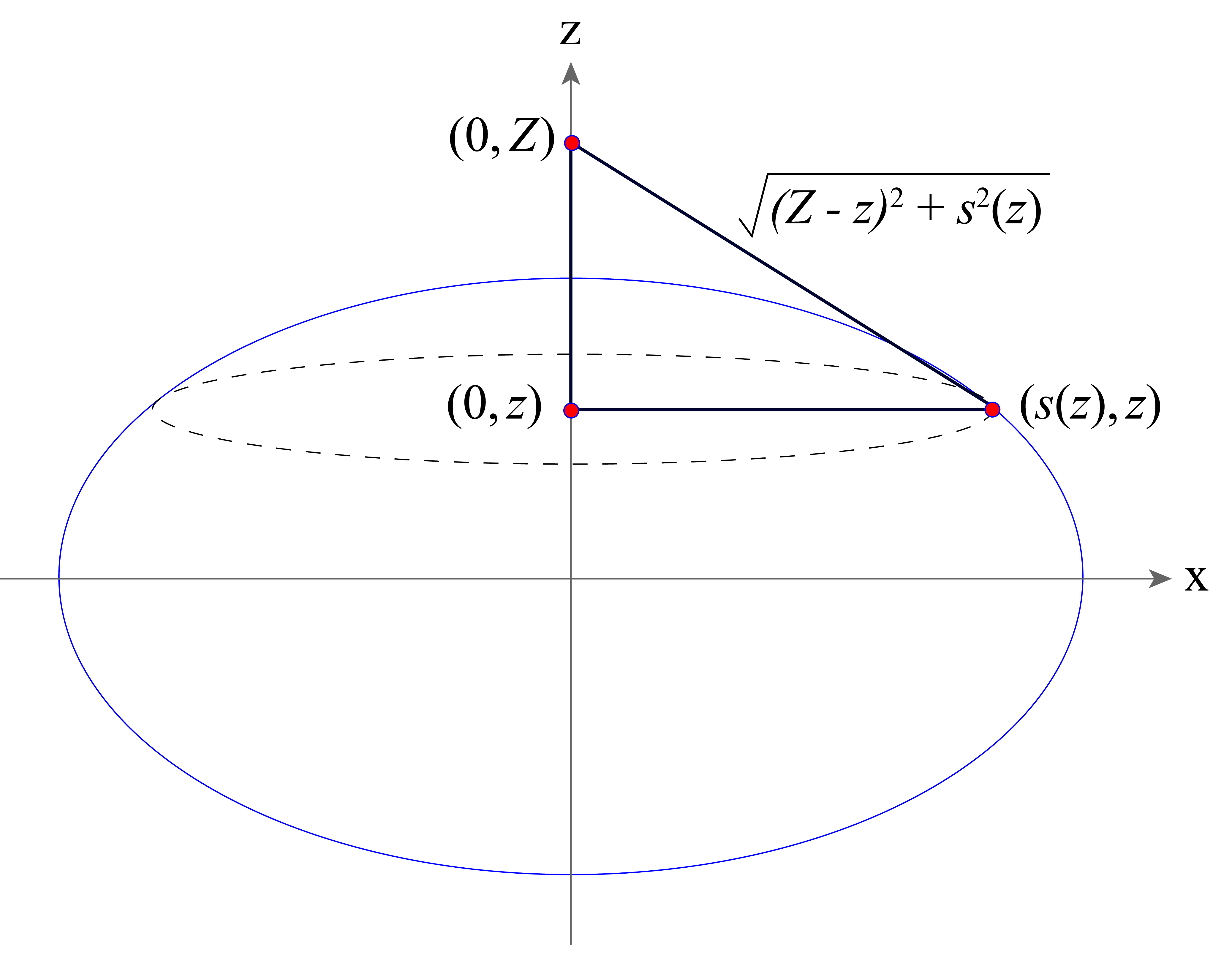}}
\caption{The blue curve shows the surface boundary curve $(\pm s(z), z)$ whose rotation about the axis of symmetry (the $z$ axis) defines the surface of the axisymmetric planet, here an oblate spheroid. The distance from the observation point, at $Z$ on the $z$ axis, to a point $(\pm s(z), z)$ on the surface curve is $\sqrt{(Z-z)^2+s^2(z)}$.}
\label{fig:axi}
\end{figure}
Here $A_n$ are numerical coefficients, $\theta$ is the angle to the axis of symmetry (the $z$ axis), and $P_n(\cos\theta)$ are the Legendre polynomials [\href{https://dlmf.nist.gov/18.12.E4}{dlmf.nist.gov/14}].\footnote{At various points in this paper, for convenience of the reader we refer to the NIST Digital Library of Mathematical Functions [\href{https://dlmf.nist.gov}{dlmf.nist.gov}] for relevant mathematical identities.}
Since the coefficients $A_n$ are independent of $\theta$, they can be computed by evaluating $\Phi(r, \theta)$ along the direction of any colatitude angle $\theta$. The simplest choice is $\theta=0$, so that $\cos\theta=1$, for which $P_n(1)=1$ for all $n$. Physically, this corresponds to the observation point $Z$ being on the positive $z$-axis, outside the planet. See figure \ref{fig:axi}. Once the coefficients $A_n$ are determined from the expansion about infinity along the positive $z$-axis, to obtain the SHE along any other colatitude direction $\theta$ we simply include the Legendre polynomial factor $P_n(\cos\theta)$ in the SHE \eqref{eq:axipot1}. This can be represented as
\begin{eqnarray}
\Phi_{z\, {\rm axis}}(r=Z, \theta=0)= \sum_{n=0}^\infty \frac{A_n}{Z^{n+1}} \quad \Rightarrow \quad
 \Phi(r, \theta)= \sum_{n=0}^\infty \frac{A_n}{r^{n+1}}\, P_n(\cos\theta)
\label{eq:axipot2}
\end{eqnarray}
In other words, the SHE along the $z$-axis determines the SHE for all $\theta$. 

\subsection{Singularities from the planetary topography}
\label{sec:sing}

For an axisymmetric planet of uniform density, $\rho$, the gravitational potential $\Phi_{z\, {\rm axis}}(Z)$ at an arbitrary point $Z$ on the positive $z$ axis exterior to the planet can be expressed as
\begin{eqnarray}
\Phi_{z\, {\rm axis}}(Z)\equiv \Phi(Z, 0) &=& 
- 2\pi G \rho \int_{z_{\rm min}}^{z_{\rm max}} dz \int_0^{s(z)} s\, ds \frac{1}{\sqrt{(Z-z)^2+s^2}}
\label{eq:pot1}
\end{eqnarray}
where $G$ is Newton's constant and $\rho$ is the uniform density.
This expression effectively integrates over each horizontal circular cross-sectional slice of radius $s(z)$, from the south to north poles of the axisymmetric plane at
$z_{\rm min}$ and $z_{\rm max}$. See figure \ref{fig:axi}. The integrand is simply the inverse of the Pythagorean distance from the observation point $(0, Z)$ to the point $(s, z)$ in the vertical cross-sectional $(x, z)$ plane. The radial variable $s$ is integrated from $0$ to the the edge of each horizontal slice at $s=s(z)$.
This  $s$ integral is elementary,  so we can write
\begin{eqnarray}
\Phi_{z\, {\rm axis}}(Z)
=
- 2\pi G \rho \int_{z_{\rm min}}^{z_{\rm max}} dz \left[\sqrt{(Z-z)^2+s^2(z)}-(Z-z)\right]
\label{eq:pot1b}
\end{eqnarray}
Therefore, the potential along the $z$ axis can be expanded in the limit $Z\to +\infty$, yielding a simple integral representation of the SHE coefficients \cite{axi}:
\begin{eqnarray}
A_n =-2\pi G \rho \int_{z_{\rm min}}^{z_{\rm max}} dz\, C_{n+2}^{(-\frac{1}{2})}\left(\frac{z}{\sqrt{z^2+s^2(z)}}\right)\left(z^2+s^2(z)\right)^{\frac{(n+2)}{2}}
\label{eq:an}
\end{eqnarray}
Here $C_{k}^{(\nu)}$ is a Gegenbauer (ultraspherical) polynomial [\href{https://dlmf.nist.gov/18.12.E4}{dlmf.nist.gov/18.12.E4}], arising from the expansion of the square root integrand in \eqref{eq:pot1b}. 
The leading term, $A_0$, is universal, in terms of  the total volume $V$, or mass $M$, of the axisymmetric planet:
\begin{eqnarray}
A_0 = -  G \rho \left(\pi \int_{z_{\rm min}}^{z_{\rm max}} dz \, s^2(z) \right) \equiv - G\rho V =-G\, M
\label{eq:a0}
\end{eqnarray}
The expression \eqref{eq:an} demonstrates that the SHE coefficients $A_n$ are completely determined by the boundary curve $s(z)$. In some special cases (e.g., the sphere, spheroid and cylinder, below) the $A_n$ can be computed analytically, and in other cases numerically. 

In this paper we present a variety of illustrative examples which highlight important  features of the use of Pad\'e approximants for truncated series of the SHE of the gravitational potential. 
We concentrate on axisymmetric examples, in order to compare the Pad\'e results with the other complementary singularity methods in \cite{axi}. 
In particular,  in \cite{axi} it is shown that 
for axisymmetric planets of uniform density, the radius of convergence of the SHE, and indeed the detailed large order ($n\to +\infty)$ behavior of the SHE coefficients $A_n$, is completely determined by the boundary curve function $s(z)$ that describes the planetary topography. This is because the convergence behavior is determined by the complex singularities of the gravitational potential, and these singularities can be found directly from $s(z)$ without computing the SHE coefficients $A_n$.
We show how this rich singularity structure arises in the Pad\'e approach, and how the gravitational potential may be analytically continued beyond the region of convergence.

In \cite{axi} it is shown that the general integral representation \eqref{eq:pot1b} for the gravitational potential implies that singularities as a function of $Z$ occur when the function
\begin{eqnarray}
p(z):=(Z-z)^2+s^2(z)
\label{eq:p}
\end{eqnarray} 
satisfies the following conditions 
\begin{eqnarray}
p(z)=0 \quad &\Rightarrow& \quad Z=z\pm i\, s(z)
\nonumber\\
p'(z)=0 \quad &\Rightarrow& \quad Z=z+s'(z)\, s(z)
\label{eq:conditions}
\end{eqnarray}
These conditions combine to the following:
\begin{eqnarray}
Z_0=z_0 \pm is(z_0)\qquad {\rm where}\qquad s'(z_0)=\pm i
\label{eq:combined}
\end{eqnarray}
Examples of this kind of singularity are discussed in sections \ref{sec:spheroid}, \ref{sec:lozenge}, \ref{sec:peanut} and \ref{sec:rough}.

Alternatively, if the boundary profile function $s(z)$ itself has singularities, which correspond to curvature singularities of the topography \cite{axi}, then these also lead to singularities of the gravitational potential, on the planet surface. 
An example of this kind of singularity is in the discussion of a cylindrical planet, in  section \ref{sec:cylinder}.

When $s(z)^2$ is a polynomial (which implies that $p(z)$ is also a polynomial), the condition in \eqref{eq:conditions} can be stated as the geometric condition that the {\it discriminant} of $p(z)$ has a root \cite{axi}. For a quadratic polynomial, $p(z)=a\, z^2+b \, z+c$, the discriminant is ${\rm Disc}[p(z)]=b^2-a c$. The vanishing of the discriminant means that the two roots of $p(z)$ are equal, which in turn means that $p(z)$ and $p'(z)$ share the same root. For higher order polynomials the {\it discriminant} ${\rm Disc}[p(z)]$ is a function of the coefficients of $p(z)$, so for $p(z)$ defined in \eqref{eq:p} the discriminant is a polynomial in $Z$, with coefficients expressed in terms of the coefficients of $s(z)^2$.
The vanishing of the discriminant is equivalent to the vanishing of both $p(z)$ and $p'(z)$, which is the condition in \eqref{eq:conditions}.

There are efficient algorithms for evaluating the discriminant of the polynomial $p(z)$. For example,  it is a built-in function in Mathematica and Maple. There are also accurate root finding methods, so high precision determination of the discriminant zeros, $Z_0$, is extremely efficient, even for very high order polynomials for $s^2(z)$. See section \ref{sec:rough}. 

\subsection{Geometric Interpretation of Complex Singularities}
\label{sec:geometric}

There is a useful geometric interpretation of the complex singularities of the SHE {\it along the axis of symmetry}, obtained by rotating the $Z$ plane by $\pi/2$. This is achieved by multiplying the complex singularities $Z_0$ in \eqref{eq:combined} by $i$:
\begin{eqnarray}
    i\, Z_0= \pm s(z_0) + i\, z_0
    \label{eq:rotated}
\end{eqnarray}
Recall from figure \ref{fig:axi} that if $z_0$ is real with $z_0\in [z_{\rm min}, z_{\rm max}]$, then a point on the planet surface is represented in the "vertical cross-section" $(x, z)$ plane as $(\pm s(z_0), z_0)$, which we can express as the complex number $\pm s(z_0)+i\, z_0$. 
For more general complex $Z_0$, we can identify the real part with the $x$-axis and its imaginary part with the $z$-axis in figure \ref{fig:axi}. 

With this identification we find the following geometric interpretation of complex singularities of the gravitational potential along the axis of symmetry:
\begin{enumerate}
    \item If $i\, Z_0$ lies on the planetary surface curve, then it is a singularity due to surface curvature, such as a vertex or an edge in a synthetic planet. This is the case for the cylinder in section \ref{sec:cylinder}.
    \item If $i\, Z_0$ lies {\it inside} the planetary surface curve, then this singularity is associated with the divergence of the SHE. The radius of convergence is determined by $|i\, Z_0|=|Z_0|$ for the interior singularity (or singularities) closest to the planetary surface. This kind of singularity is discussed in sections \ref{sec:spheroid}, \ref{sec:lozenge}, \ref{sec:peanut} and \ref{sec:rough}.
    \item
    If $i\, Z_0$ lies {\it outside} the planetary surface curve, then this singularity is associated with the next Riemann sheet and does not affect the divergence of the SHE. This kind of singularity is discussed in sections \ref{sec:lozenge}, \ref{sec:peanut} and \ref{sec:rough}.
\end{enumerate}

\subsection{Pad\'e Approximants: Convergence and Potential-Theoretic Interpretation}
\label{sec:pade}

We summarize here several results on the convergence of Pad\'e approximants and their connection to potential theory, following \cite{stahl,pade-noise,cmp-pade,epj-pade}, to which we refer for proofs and further details.  See also \S\ref{S:Branch} below.

Let $f$ be a function analytic at infinity with finitely many  {\bf branch point singularities}  in $\mathbb{C}$ (see section \ref{S:Branch}). Denote by $\mathcal{D}$ a domain of single-valuedness of $f$, connecting the singularities so that any closed path in $\mathcal{D}$ cannot wind around any singularity, and let $E=\partial \mathcal{D}$ be the corresponding boundary.   The set $E$ consists of a system of piecewise analytic arcs joining branch points, possibly with additional junction points. For concrete examples see the red dots in figures \ref{fig:pade-spheroid}, \ref{fig:cylinder},  \ref{fig:pade-cylinder-noise}, \ref{fig:lozenge1}, \ref{fig:peanut-pade} and \ref{fig:cheb-pade}. In the limit that the order of Pad\'e increases, these dots form continuous piecewise analytic arcs that make up the boundary $E=\partial \mathcal{D}$ of the region of single-valuedness. 

A fundamental result due to Stahl \cite{stahl} identifies a distinguished set $E$, characterized as the curve (continuum, more precisely) of minimal logarithmic capacity among all such admissible branch cut configurations. Intuitively, these cuts may be viewed as pieces of an infinitely thin, flexible wire forming a connected conductor. One may think of this conductor as a two-dimensional capacitor, whose shape is allowed to vary (while still connecting the singularities) so as to minimize its capacity with respect to infinity. The resulting {\it minimal capacitor} forms  the boundary of the region of single-valuedness produced by Pad\'e in the large order limit.
This is somewhat analogous to a one-dimensional version of  Plateau's problem, the minimization of soap bubbles \cite{douglas}.

The complement $\mathcal{D}=\mathbb{C}\setminus E$ is then the {\it domain of convergence} (see below) of near-diagonal Pad\'e approximants to $f$.

This construction admits a natural interpretation in logarithmic potential theory \cite{stahl}. Viewing $E$ as a conductor carrying a unit charge, the equilibrium configuration minimizes the discrete energy
\[
\mathcal{E}_N = -\sum_{1\le i<j\le N} \log|\omega_i - \omega_j|,
\]
whose minimizers are the Fekete points, representing the equilibrium positions of discrete charges. Recall that electrical charges interact by a logarithmic potential in two dimensions \cite{stahl,cmp-pade,epj-pade}.  As $N\to\infty$, these configurations converge to the equilibrium measure on $E$, and the poles of near-diagonal Pad\'e approximants accumulate along $E$ according to this distribution.

Within the domain $\mathcal{D}$, the rate of convergence of Pad\'e approximants is governed by the Green's function $G_{\mathcal{D}}$ with pole at infinity. More precisely, for $\omega\in \mathcal{D}$,the deviation of the $2N$-term Taylor series $f(\omega)$ from its diagonal Pad\'e approximant $[N,N]_f(\omega)$ is given by 
\[
|f(\omega)-[N,N]_f(\omega)|^{1/2N} \to |\psi(\omega)| \quad, \quad  \text{as } N\to \infty
\]
where $\psi$ is the Riemann conformal map (analytic map, together with its inverse) from $\mathcal{D}$ onto the unit disk, normalized at infinity. Thus, Pad\'e approximants encode the conformal geometry of the maximal domain of analyticity. Physically, this is because two dimensional electrostatics problems are solved by conformal maps.

The convergence is understood in capacity (i.e., except possibly for sets of zero logarithmic capacity, such as isolated points). This also explains the appearance of spurious poles, which can be removed to recover uniform convergence on compact subsets of $\mathcal{D}$.

These results show that Pad\'e approximants effectively reconstruct both the domain of analyticity and the associated conformal map, providing a natural framework for analytic continuation {\it beyond the radius of convergence of the original series}.

\subsection{Physically relevant singularities vs.\ ``higher-sheet'' ones. Branch points.}\label{S:Branch}

Many singularities arising in synthetic planetary models---for instance in formulas of the {\it Werner--Scheeres} type \cite{werner} ---are {\it branch points} \cite{fokas}. The simplest example is the square root $g(z)=\sqrt{z}$, which is singular at $z=0$. Writing $z=|z|e^{i\varphi}$ with $z=x+iy$, $|z|=\sqrt{x^2+y^2}$ and $\varphi=\arg z$ ($\tan\varphi=y/x$), we have
\[
\sqrt{z}=\sqrt{|z|}\,e^{i\varphi/2}.
\]
If we start at $z=1$ (so $\varphi=0$), then $\sqrt{z}=1$. After one full rotation around the origin, $\varphi\mapsto\varphi+2\pi$, we return to the same point $z=1$ but obtain
\[
\sqrt{z} \text{\,\, becomes\,\, } \sqrt{|1|}\, e^{i(\varphi+2\pi)/2}=-1.
\]
Thus the value of the function changes after circling the point $z=0$. Such functions are called {\it multi-valued}, and the singularity at $z=0$ is a {\it branch point}.

This contrasts with $f(z)=1/z$, which is also singular at $0$ but satisfies
\[
\frac{1}{z}=\frac{1}{|z|}e^{-i\varphi},
\]
so that after a full rotation one returns to the same value. Similarly, functions such as polynomials and rational functions are also {\it single-valued}. In particular, Pad\'e approximants, being rational functions, are single-valued.

Because of this, a rational function cannot reproduce the behavior of $\sqrt{z}$ on any region that allows a full loop around $0$: the function $\sqrt{z}$ changes sign, while a rational function does not. One introduces {\it branch cuts} that prevent such loops. As discussed in section \ref{sec:pade}, for Pad\'e approximants, these are reflected in the accumulation of poles along curves approximating cuts.

Branch points can also lead to subtler effects. Consider
\[
h(z)=\frac{1-\sqrt{1-z}}{z}.
\]
Although the formula appears singular at $z=0$, this is only a {\it removable singularity}. Expanding near $z=0$ gives
\[
h(z)=\frac{1}{2}+\frac{z}{8}+\frac{z^2}{16}+\frac{5z^3}{128}+\frac{7z^4}{256}+\cdots,
\]
which converges for $|z|<1$, the radius being determined by the branch point at $z=1$.

However, if one analytically continues $h$ around $z=1$, the square root changes sign and one obtains a different branch,
\[
h_1(z)=\frac{1+\sqrt{1-z}}{z}=\frac{2}{z}-\frac{1}{2}-\frac{z}{8}-\frac{z^2}{16}-\frac{5z^3}{128}-\frac{7z^4}{256}+\cdots,
\]
which now has a genuine singularity at $z=0$ and no Taylor expansion there. We say that these singularities lie on {\it other Riemann sheets}.

For our purposes, the key point is the following: {\it only singularities that can be reached without crossing a branch cut determine the convergence of Taylor series and of Pad\'e approximants}. Singularities that arise only after analytic continuation around a branch point do not influence convergence. Accordingly, Pad\'e approximants detect only the singularities that are relevant for convergence. Examples of this basic phenomenon in the Pad\'e approach to geodesy are shown in section \ref{sec:examples} below. 

In Figures \ref{fig:lozenge1} and \ref{fig:peanut-pade} singularities marked by black dots that are not approached by Pad\'e poles lie on other Riemann sheets and are not relevant for convergence.

\section{Pad\'e Singularities and Discriminant  Singularities}
\label{sec:examples}

In this section we present a collection of illustrative examples deriving the SHE singularities found numerically as poles of a Pad\'e approximant to the gravitational potential, and comparing with results from the other approaches developed in \cite{axi}.
Specifically, we compute  the poles of the $[N, N]$ Pad\'e approximant  to the truncated large $Z$ expansion of the gravitational potential:
\begin{eqnarray}
    \Phi_{2N}(Z, 0):=\frac{1}{Z}\sum_{n=0}^{2N} \frac{A_n}{Z^n}
    \quad \longrightarrow \quad
    \Phi^{\rm Pade}_{[N,N]}(Z, 0)=
    \frac{1}{Z}\frac{Q(Z)}{R(Z)}
    &=&\frac{1}{Z}\frac{\sum_{n=0}^N 
    B_n/Z^n}{\sum_{n=0}^N C_n/Z^n}
    \nonumber\\
    &=& \frac{1}{Z}\sum_{n=0}^{2N} \frac{A_n}{Z^n} + O\left(\frac{1}{Z^{2N+2}}\right)
\end{eqnarray}
We compute the SHE coefficients $A_n$ using \eqref{eq:an}, either analytically or numerically.
We then find the roots $Z_0$ of the Pad\'e denominator polynomial $R(Z)$, and normalize them as $i\, Z_0$ in \eqref{eq:rotated} in order to compare with the planetary topography, as discussed in section \ref{sec:geometric}.

\begin{enumerate}
    \item The main result is that the Pad\'e poles accumulate to the roots of the discriminant that lie inside the boundary curve; these accumulation points are the singularities on the first Riemann sheet, which govern the convergence properties of the SHE.
    \item Furthermore, the Pad\'e approximant gives an  analytic continuation of the gravitational potential beyond its radius of convergence.
    \end{enumerate}

\subsection{Oblate Spheroidal Planet}
\label{sec:spheroid}

A spheroidal planet has boundary curve $s(z)=a \sqrt{1-z^2/b^2}$, with $z\in [-b, b]$. For an oblate spheroid, $a>b$, while  $b>a$ for a prolate spheroid. For definiteness, we consider an oblate spheroid with constant density.
The SHE coefficients in \eqref{eq:an} can be evaluated in closed form \cite{axi}:
\begin{eqnarray}
A_{2n}^{\rm spheroid}
&=&-3 G M \frac{(-1)^n (a^2 - b^2)^n}{(2 n + 1)(2 n + 3)}
\label{eq:an-ellipsoid}
\end{eqnarray}
Here $G$ is Newton's constant and $M$ is the planetary mass. 
A simple ratio test determines the radius of convergence of the SHE for this oblate spheroidal planet to be 
\begin{eqnarray}
R^{\rm spheroid} =\sqrt{a^2-b^2}
\label{eq:spheroidal-radius}
\end{eqnarray}
\begin{figure}[h!]
\centerline{
\includegraphics[scale=0.5]{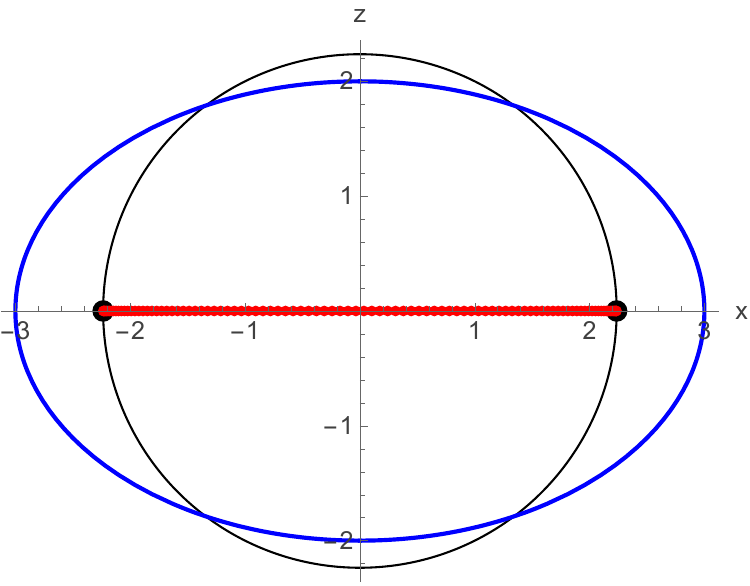}}
\caption{The rotated Pad\'e poles $i\, Z_0$ (red points) for the SHE for an oblate spheroid, along the symmetry axis (the $z$ axis). The minor axis has length 2 and the major axis has length 3. The bounding curve of the spheroid cross-section is shown in blue. The Pad\'e poles accumulate to branch points (black dots) at the locations $\pm \sqrt{5}$ of the foci of the spheroid. The radius of convergence is shown as a black circle centered at the center of mass and passing through these branch points.}
\label{fig:pade-spheroid}
\end{figure}
This can also be seen by summing the SHE along the $z$ axis:
\begin{eqnarray}
\Phi^{\rm spheroid}_{z\, {\rm axis}}(Z)=\frac{G M}{Z} \frac{3}{2}\left[\frac{Z^2}{(a^2-b^2)} +i\,\frac{Z(Z^2+(a^2-b^2))}{2(a^2-b^2)^{3/2}} \log\left(\frac{Z+i\sqrt{a^2-b^2}}{Z-i\sqrt{a^2-b^2}}\right)\right]
\label{eq:ellipsoid-final}
\end{eqnarray}
This expression identifies the  branch point singularities (rotated by $i$, as in \eqref{eq:rotated}) as 
\begin{eqnarray}
   i\, Z_0=  \pm \sqrt{a^2-b^2}
\end{eqnarray}
consistent with the radius of convergence in \eqref{eq:spheroidal-radius}. 
This is also consistent with the roots of the discriminant \cite{axi}:
\begin{eqnarray}
    {\rm Disc}\left[(Z-z)^2+s(z)^2\right]=\frac{4 a^2}{b^2} \left(Z^2+(a^2-b^2)\right)
\end{eqnarray}
To illustrate the simplicity of the Pad\'e approach, we make a diagonal Pad\'e approximation of the gravitational potential expanded about infinity along the $z$ axis. The  poles of this Pad\'e approximant are shown as red dots in figure \ref{fig:pade-spheroid}. The Pad\'e poles accumulate to branch points at the foci of the spheroid. This means that if we expand about infinity along the $z$ axis, we can analytically continue well beyond the radius of convergence, which is set by the black circle in figure \ref{fig:pade-spheroid}.

\subsection{Cylindrical Planet and Pad\'e downward continuation down to the topography}
\label{sec:cylinder}

\begin{figure}[h!]
\centerline{\includegraphics[scale=0.5]{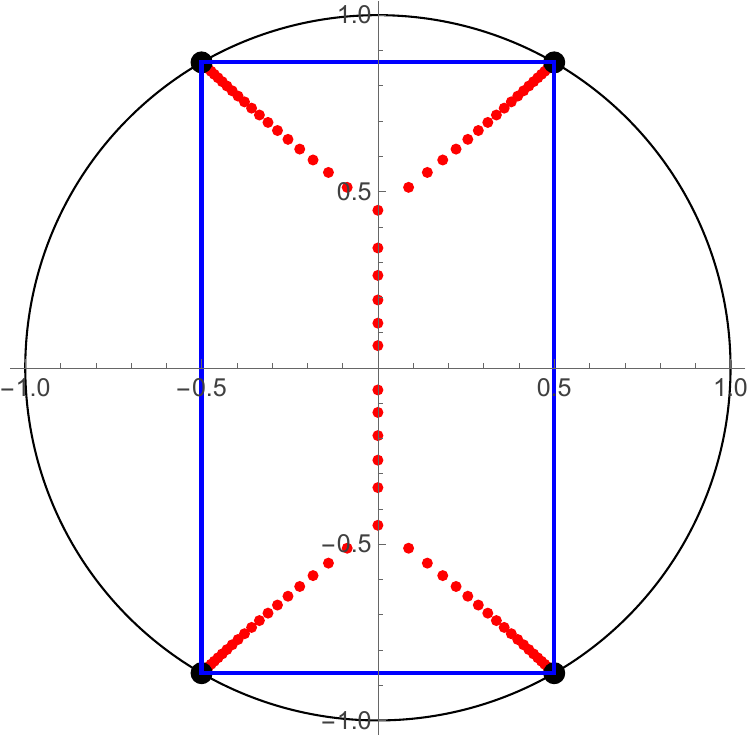}}
\caption{The blue lines show the $(x, z)$ plane cross-section of the cylindrical planet with shape function $s(z)=a=1/2$ and height $L=\sqrt{3}$. The black points show the complex singularities $i\, Z_0$ associated with the curvature singularities at the corners of the rectangular cross-section. The red dots show the Pad\'e poles (multiplied by $i$ in order to fit with the geometric picture in section \ref{sec:geometric}). 
The black circle shows the radius of convergence, here the radius of the Brillouin sphere. Note that the Pad\'e poles, $i\, Z_0$, accumulate to branch points at the four curvature singularities. }
\label{fig:cylinder}
\end{figure}
A cylindrical planet of length $L$ and radius $a$ is described by constant $s(z)=a$, with $z\in [-L/2,+L/2]$. See the blue edges in figure \ref{fig:cylinder}. The cross-sectional shape has sharp corners in the $(x,z)$ plane at $(x,z) =\left(\pm a, \pm \frac{L}{2}\right)$ and $\left(\pm a, \mp \frac{L}{2}\right)$. Therefore, from the discussion in section \ref{sec:geometric}, we expect the radius of convergence to be determined by these curvature singularities on the planetary surface.
This radius of convergence gives the radius of the Brillouin sphere, shown as a black circle in figure \ref{fig:cylinder}. As described in section \ref{sec:geometric}, these curvature singularities are identified with the rotated complex singularities $i\, Z_0=\pm a \mp i\, L/2$.

The coefficients of the SHE along the axis of symmetry, expanded at infinity on the positive $z$ axis, can be evaluated in closed form \cite{axi} (the odd-indexed coefficients vanish, by symmetry): 
\begin{eqnarray}
A_{2n}^{\rm cylinder}
&=&- G M a^{2n} \frac{\left(\frac{4 a}{L} \right)\left(1+\left(\frac{L}{2a}\right)^2\right)^{n+1/2}}{(2n+1)(2n+2)(2n+3)}\,
C_{2n+1}^{(3/2)}\left(\frac{L/(2a)}{\sqrt{1+\left(\frac{L}{2a}\right)^2}}\right)
\label{eq:an-cylinder}
\end{eqnarray}
Here $C_n^{(\nu)}$ is a Gegenbauer (ultraspherical) polynomial [\href{https://dlmf.nist.gov/18.12.E4}{dlmf.nist.gov/18.12.E4}].
For given $(a, L)$ parameters for the radius $a$ and length $L$ of the cylinder, it is straightforward to generate many of these coefficients.
We use these coefficients to make a diagonal Pad\'e approximant for the gravitational potential along the axis of symmetry, from which it is straightforward to find the Pad\'e poles. These are shown as red dots in figure \ref{fig:cylinder}. In the limit of large Pad\'e order, these red dots form the arcs of the {\it minimal capacitor} described in section \ref{sec:pade}. The Pad\'e approximant in this figure was made starting with 100 terms of the SHE for the cylinder in \eqref{eq:an-cylinder}, with parameters $a=1/2$ and $L=\sqrt{3}$. Therefore the Brillouin sphere has radius $\sqrt{a^2+(L/2)^2}=1$.
\begin{figure}[h!]
\centerline{\includegraphics[scale=.3]{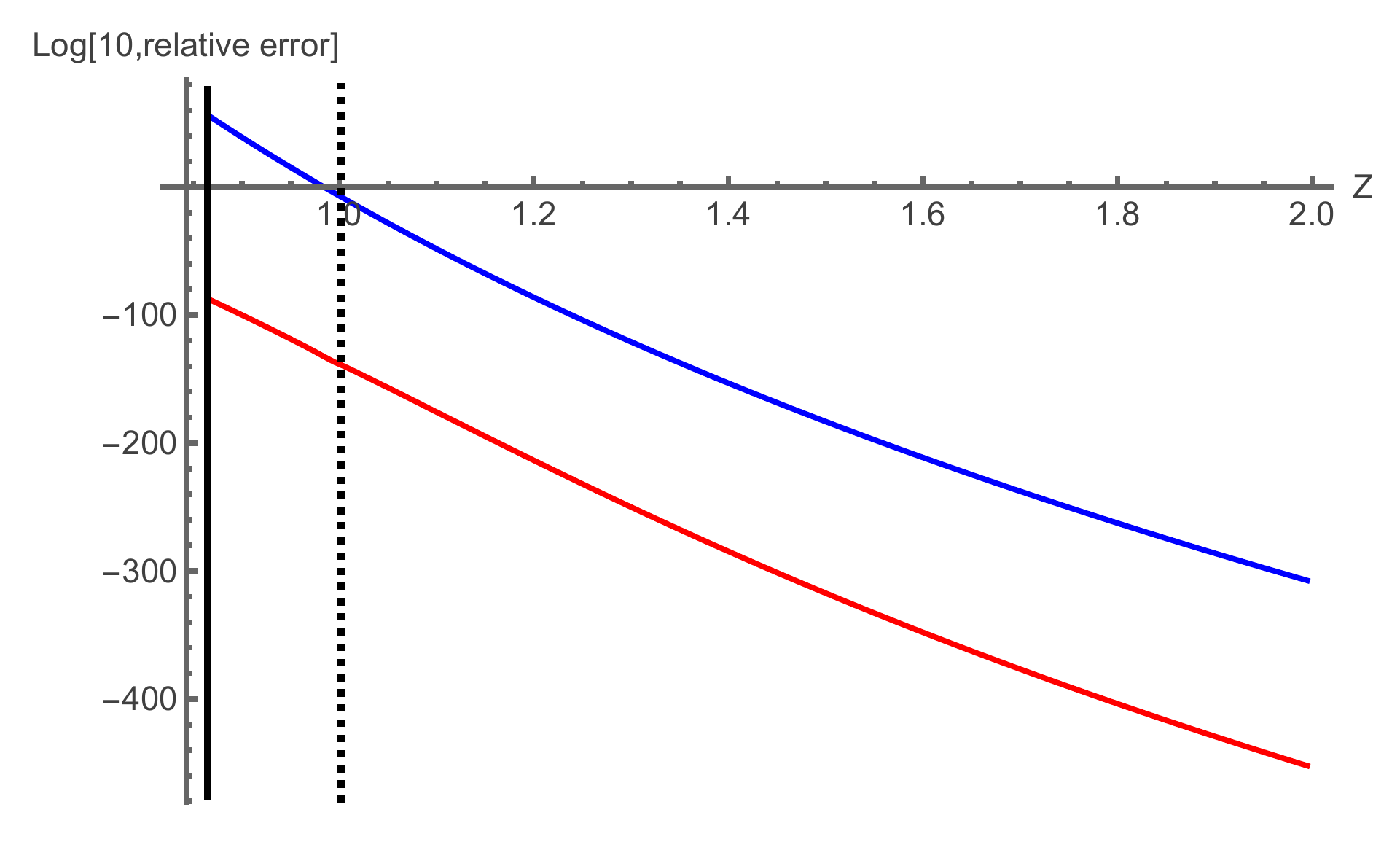}}
\caption{This plot shows the base 10 logarithm of the relative error of the 500 term SHE [blue] for the cylindrical planet with $(a, L)=(1/2,\sqrt{3})$ and the corresponding diagonal Pad\'e approximant [red]. The Brillouin radius is at $Z=1$, denoted by the vertical dashed line, and the planetary boundary is denoted by the vertical solid line at $Z=\sqrt{3}/2$. The Pad\'e approximant is dramatically more accurate, especially near and below the Brillouin radius, all the way down to the planetary surface. }
\label{fig:cylinder-down}
\end{figure}

The Pad\'e approximant also provides an analytic continuation of the gravitational potential along the axis of symmetry that goes further towards the planet than the  Brillouin sphere. In fact, the Pad\'e approximant along the axis of symmetry continues smoothly all the way to the top cap of the cylinder, at $z=\sqrt{3}/2=0.866$. The base 10 logarithm of the relative error of this downward continuation of the 500-term Pad\'e approximant is shown by the red curve in figure \ref{fig:cylinder-down}, in contrast to the relative error of the 500-term SHE itself, shown in blue. The Pad\'e approximant is dramatically more precise than the SHE, especially below the Brillouin sphere radius $Z=1$, all the way to the surface of the planet at $Z=\sqrt{3}/{2}$.  

{\bf Note}. This downward continuation feature holds for planets of arbitrary shape and will be discussed in a forthcoming paper:  for sufficiently high Pad\'e order and numerical accuracy of the coefficients, Pad\'e approximants are valid down to an arbitrarily small distance from {\it the surface of the planet}.

\begin{figure}[h!]
\centerline{\includegraphics[scale=0.5]{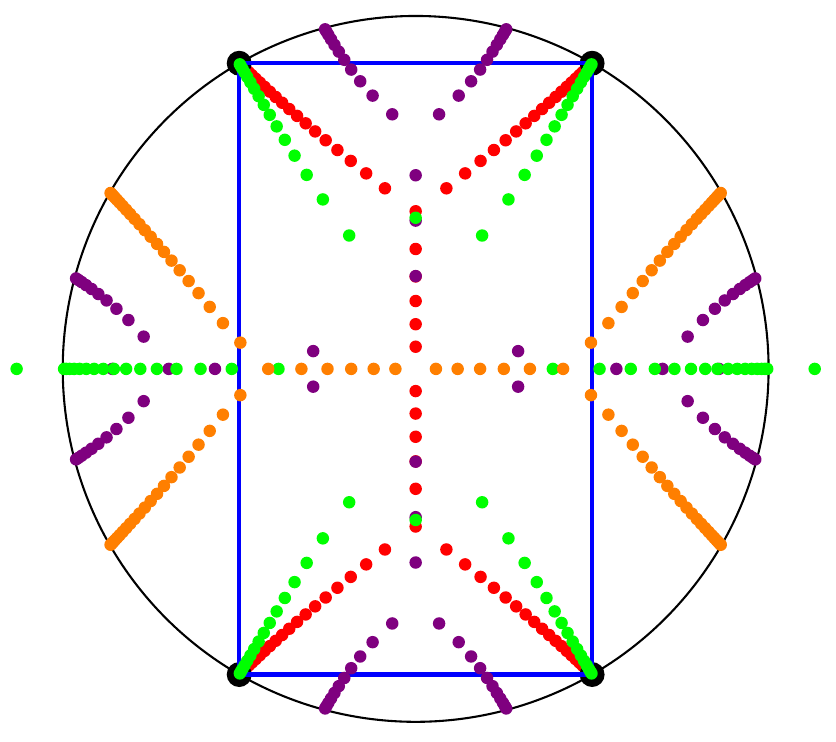}}
\caption{The blue lines show the $(x, z)$ plane cross-section of the cylindrical planet with shape function $s(z)=a=1/2$ and height $L=\sqrt{3}$. The black points show the complex singularities $i\, Z_0$ associated with the curvature singularities at the corners of the rectangular cross-section. The colored dots show the Pad\'e poles (multiplied by $i$ in order to fit with the geometric picture in section \ref{sec:geometric}) for the SHE along various colatitude directions: $\theta=0$ (red); $\theta=\pi/4$ (purple); $\theta=\pi/3$ (green); $\theta=\pi/2$ (orange).
The black circle shows the radius of convergence, here the radius of the Brillouin sphere. }
\label{fig:multi-cylinder-pade}
\end{figure}
To study the convergence of the SHE along a direction with colatitude angle $\theta$, we simply multiply the SHE coefficients in \eqref{eq:an-cylinder} by a factor of $P_{2n}(\cos\theta)$. 
\begin{eqnarray}
A_{2n}^{\rm cylinder}(\theta)
&=&- G M a^{2n} \frac{\left(\frac{4 a}{L} \right)\left(1+\left(\frac{L}{2a}\right)^2\right)^{n+1/2}}{(2n+1)(2n+2)(2n+3)}\,
C_{2n+1}^{(3/2)}\left(\frac{L/(2a)}{\sqrt{1+\left(\frac{L}{2a}\right)^2}}\right)P_{2n}(\cos\theta)
\label{eq:an-cylinder-theta}
\end{eqnarray}
We can then make a Pad\'e approximant to this truncated series, now in inverse powers of $r$, the radial distance along the colatitude direction $\theta$. The resulting Pad\'e poles (each multiplied by a factor of $i$) are shown in figure \ref{fig:multi-cylinder-pade}. Note that the Pad\'e poles, $i\, Z_0$, accumulate to branch points on the Brillouin sphere, but rotated at different angles. This intricate $\theta$ dependence will be discussed further in a forthcoming paper. 
\begin{figure}[h!]
\centerline{\includegraphics[scale=0.5]{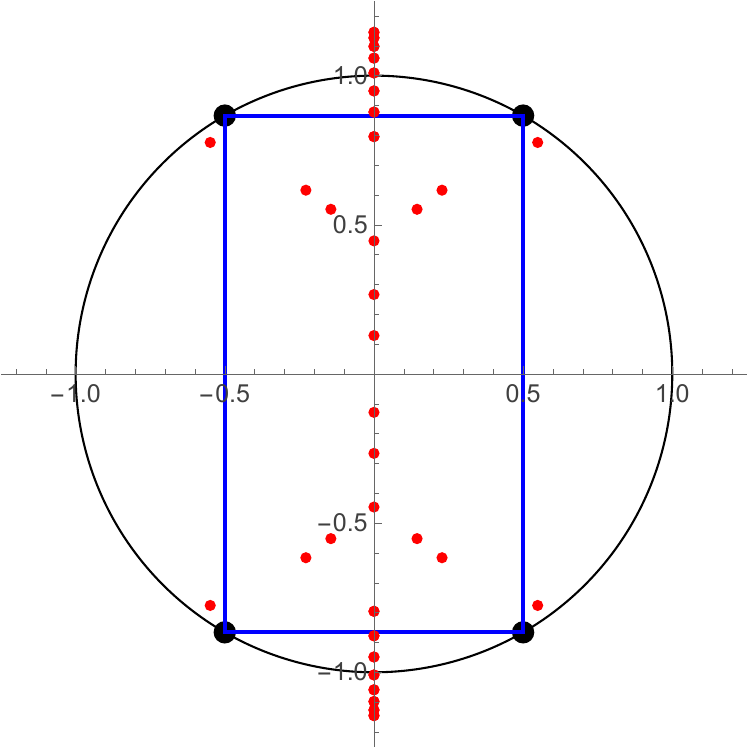}\quad 
\includegraphics[scale=0.5]{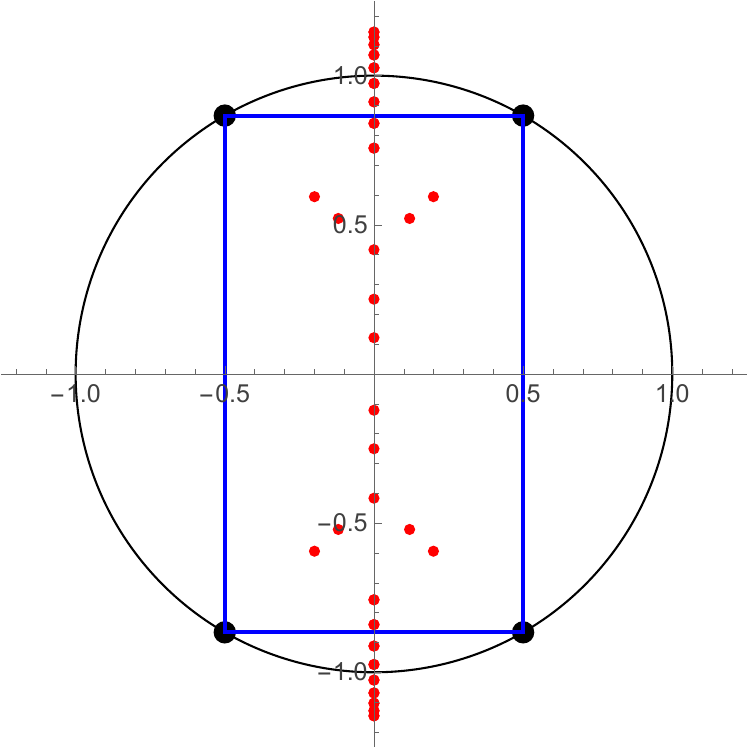}}
\centerline{\includegraphics[scale=0.5]{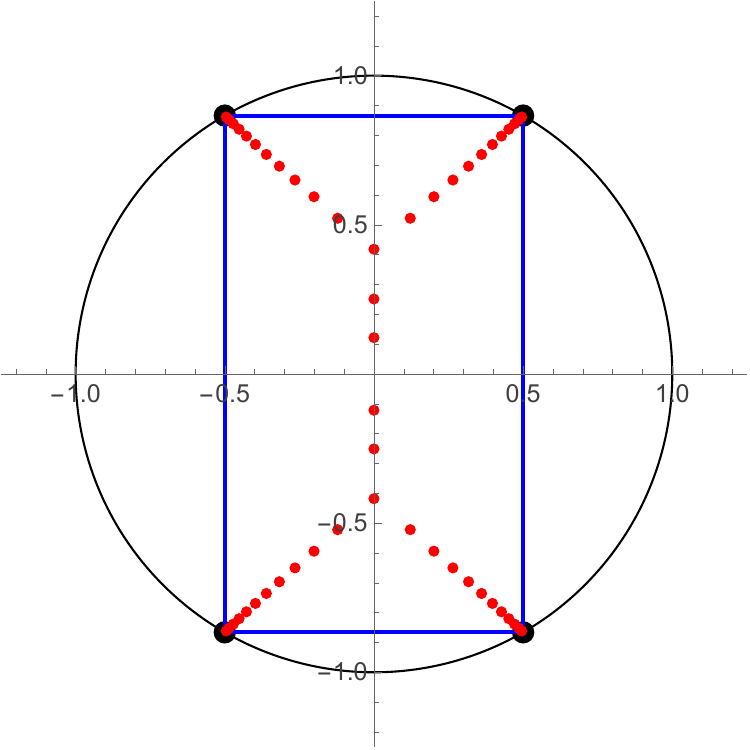}\quad 
\includegraphics[scale=0.5]{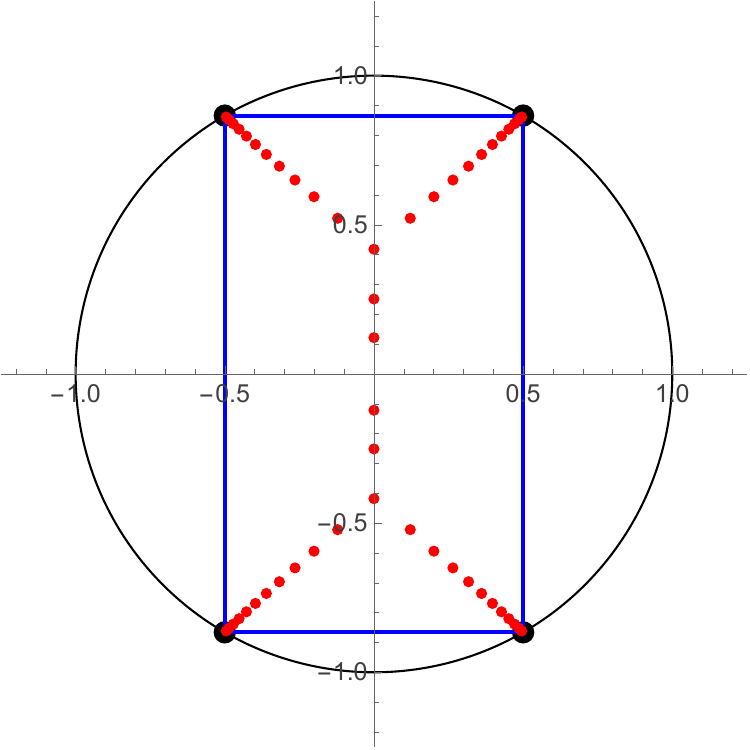}}
\caption{This figure illustrates the effect on the Pad\'e analysis of SHE coefficients with different levels of numerical precision. These plots show the Pad\'e poles, $i\, Z_0$, for the cylinder with $a=1/2$ and $L=\sqrt{3}$, with 50 SHE coefficients given by \eqref{eq:an-cylinder}, but restricted to $10$, $11$, $12$ and $13$ digits of precision [top left to bottom right].}
\label{fig:pade-cylinder-noise}
\end{figure}

It is also interesting to note the effect on the Pad\'e approximant method of the {\it finite precision} of the SHE coefficients. In figure \ref{fig:pade-cylinder-noise} we show the normalized Pad\'e poles, $i\, Z_0$, for the gravitational potential along the axis of symmetry, with the SHE coefficients evaluated from \eqref{eq:an-cylinder} with $10$, $11$, $12$ and $13$ digits of precision. This shows the generic feature of Pad\'e being seriously degraded when taking too many coefficients with too low precision. For a given precision there is typically a sharp threshold of the number of SHE coefficients below which Pad\'e is not reliable \cite{pade-noise}.

\begin{figure}[h!]
\centering
\includegraphics[scale=0.5]{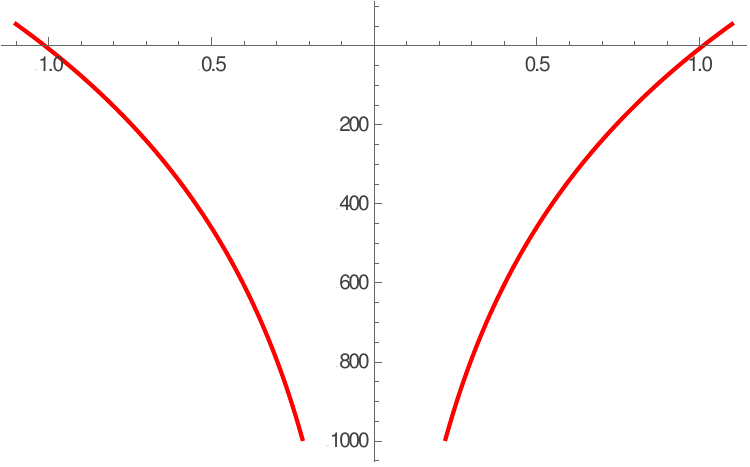}\quad
\includegraphics[scale=0.5]{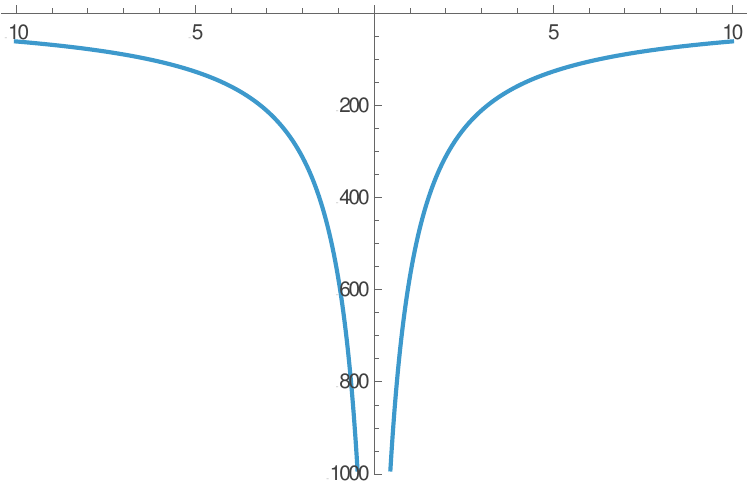}
\includegraphics[scale=0.5]{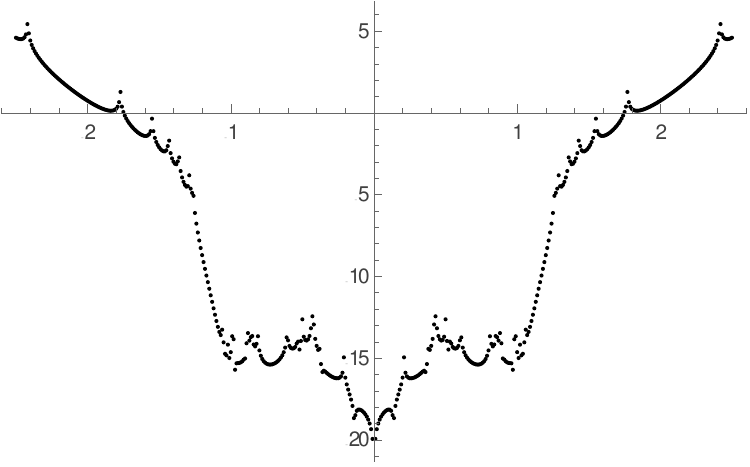}
\caption{
Comparison of the function $f(z)=(1+z^2)^{3/2}$, which has branch point singularities of order $3/2$ at $z=\pm i$ and modulus $1$, representative for the potential of non-smooth planetary topography, with its truncated series expansion (1500 terms) and Pad\'e approximants of type $[750,750]$. The Pad\'e approximants are computed both in high precision (1000 digits) and in standard machine precision (16 digits). Errors are shown as $\log_{10}$ of the absolute error, with red, blue, and black curves corresponding to the series, high-precision Pad\'e, and machine-precision Pad\'e, respectively. Thus, the negative of the vertical coordinate represents the number of correct digits. The series approximation loses accuracy outside its domain of convergence $[-1,1]$, while the high-precision Pad\'e approximant achieves approximately 10 digits of accuracy at distances up to an order of magnitude beyond this radius. Notably, the Pad\'e approximation is more accurate than the truncated series even within the interval $[-1,1]$.
Even in machine precision, the Pad\'e approximant provides limited continuation beyond the radius of convergence, with rapid deterioration near $z=\pm 1$. SHEs of planetary data are, at this time, far less accurate than machine precision.
}
\label{fig:Extrapolation}
\end{figure}

For example, in Figure \ref{fig:Extrapolation} we compare the function $f(z)=(1+z^2)^{3/2}$, which has branch point singularities of order $3/2$ at $z=\pm i$ (with $|\pm i|=1$, representative for the potential of non-smooth planetary topography \cite{bevis-2024,axi}), with its truncated series expansion (1500 terms) and Pad\'e approximants of type $[750,750]$, computed both in high precision (1000 digits) and in standard machine precision (16 digits). Spherical harmonic expansions of planetary data are typically far less accurate than machine precision. Analytic results for the  effects of noise on Pad\'e approximants are given in \cite{pade-noise}. For noisy data, more robust approaches are available; their discussion lies beyond the scope of the current paper.

\subsection{"Smoothed Cylinder" Planet}
\label{sec:lozenge}

\begin{figure}[h!]
\centerline{\includegraphics[scale=0.75]{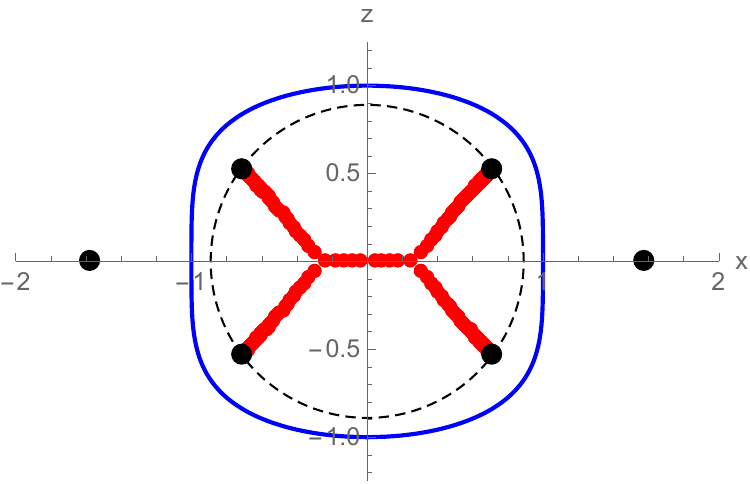}}
\caption{Boundary curve [blue] and discriminant zeros  [black dots] for the smoothed cylinder planet having $s(z)=\sqrt{1-z^4}$. The boundary of the region of  convergence of the SHE is shown by the dashed circle, whose radius is determined by the discriminant zeros lying inside the planetary cross-section.}
\label{fig:lozenge1}
\end{figure}

The boundary profile function $s(z)=\sqrt{1-z^4}$ represents a ``smoothed cylinder". See the blue curve in  Figure \ref{fig:lozenge1}. Unlike the planet in the previous section, this planet has no curvature singularities. Therefore the singularities of the gravitational potential along the axis of symmetry are determined by the discriminant in \eqref{eq:p}.
\begin{figure}[h!]
\centerline{\includegraphics[scale=0.75]{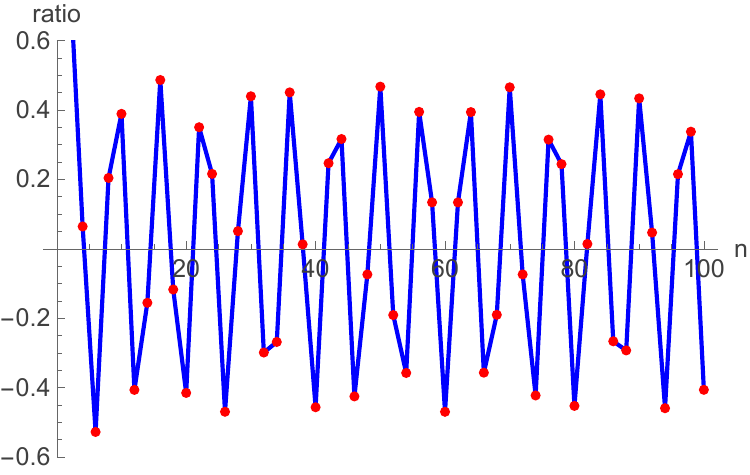}}
\caption{The red points show the exact ratio $n^2 A_{2n}/(|Z_0|)^{2n}$, for the SHE coefficients $A_{2n}$ for the smoothed cylinder planet with shape function $s(z)=\sqrt{1-z^4}$. Here $Z_0$ is one of the 4 roots of the discriminant in \eqref{eq:delta-lozenge}, such that $i Z_0$ lies inside the planetary boundary curve, as shown in Figure \ref{fig:lozenge1}. }
\label{fig:smooth-cube-ratio}
\end{figure}
For the polynomial $p(z)=(Z-z)^2+(1-z^4)$ the discriminant is 
\begin{eqnarray}
{\rm Disc}(Z)=-16 \left(16 Z^6+47 Z^4+28 Z^2+25\right)
\label{eq:delta-lozenge}
\end{eqnarray}
This is a sextic polynomial in $Z$, but since it is a cubic polynomial in $Z^2$, the roots $Z_0$ can be found explicitly. The numerical values of $i Z_0$ for these six roots are
\begin{eqnarray}
i\, Z_0= \{\pm 1.577, -0.7135 \pm 0.5325 i, 0.7135\pm 0.5325  i\}
\end{eqnarray}
These values of $i\, Z_0$ are shown as black dots in figure \ref{fig:lozenge1}. We observe that four of these roots lie inside the planet, as two complex conjugate pairs, while the two on the real axis lie outside the planet. The four interior roots are all equidistant from the origin (the center of mass).

In figure \ref{fig:lozenge1},  the red dots show the Pad\'e poles (multiplied by $i$), and we see that these accumulate precisely to the four (normalized) discriminant roots interior to the planetary surface, and not to the discriminant roots lying outside the planet.

To confirm that these Pad\'e poles determine the radius of convergence of the SHE, in figure \ref{fig:smooth-cube-ratio} we plot the ratio  $n^2 A_{n}/(|Z_0|)^{n}$, for $|Z_0|$ the magnitude of each of the four interior Pad\'e poles. The oscillatory behavior of the ratio is due to the complex conjugate pairs of singularities. The fact that the overall amplitude is not growing or decaying  shows that the radius of convergence of the SHE is determined by these interior Pad\'e poles. The discriminant roots such that $i Z_0$ lies outside the planetary boundary curve do not affect the convergence properties of the SHE.

\subsection{"Peanut" Planets}
\label{sec:peanut}

\begin{figure}[h!]
\centerline{\includegraphics[scale=.5]{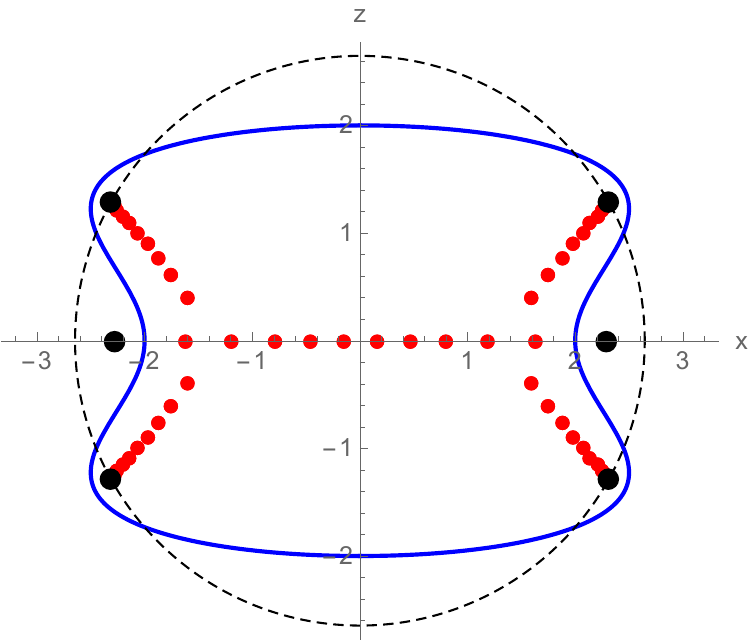}
\qquad \includegraphics[scale=.47]{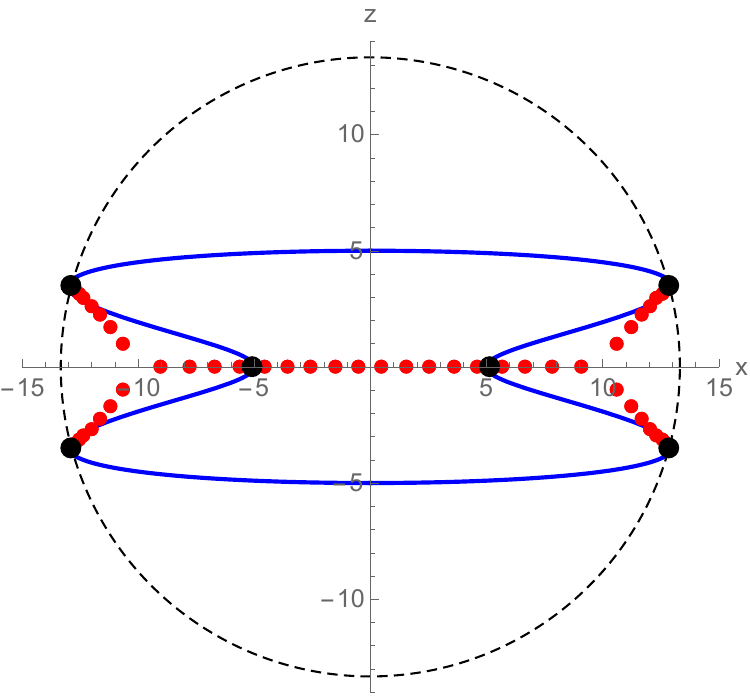}}
\caption{Pad\'e poles (red points) and roots of the discriminant (black points), for the peanut shapes: $s(z)^2=4+3z^2-z^4$ (left)
and $s(z)^2=25+24z^2-z^4$  (right). The Pad\'e poles accumulate to branch points of the gravitational potential that lie inside the planet. 
}
\label{fig:peanut-pade}
\end{figure}

In this section we analyze planets that have concave parts of the boundary surface curve. The boundary curve has regions of both positive and negative curvature. See the blue curves in figure \ref{fig:peanut-pade}, which have squared boundary shape functions:
\begin{eqnarray}
s(z)^2&=&4+3z^2-z^4 
\\
s(z)^2&=&25+24z^2-z^4 
\label{eq:peanuts}
\end{eqnarray}

The corresponding discriminants are
\begin{eqnarray}
{\rm Disc}\left[ (Z-z)^2+4+3z^2 -z^4 \right]
&=& -16 \left(16 Z^6+203 Z^4+1408 Z^2+4096\right)
\\
{\rm Disc}\left[ (Z-z)^2+25+24 z^2 - z^4 \right]
&=& -16 \left(16 Z^6+5327 Z^4+632500 Z^2+13140625\right)
\label{eq:peanut-discs}
\end{eqnarray}
The corresponding  discriminant zeros (normalized by a factor of $i$) are:
\begin{eqnarray}
D(Z)=0 &\Rightarrow&  i\,Z_0= \{\pm 2.28542, \pm 2.31658 \pm  1.27842 i \}
\\
D(Z)=0 &\Rightarrow& i\, Z_0= \{\pm 5.10293,\pm 12.8655\pm 3.47457 i\}
   \label{eq:peanuts-z0}
\end{eqnarray}

In Figure \ref{fig:peanut-pade} we show the  Pad\'e poles (red points), together with the discriminant zeros, $i\, Z_0$, (black points) for four different choices of parameters, with the corresponding boundary shape curves shown as blue curves. The radius of convergence is indicated by the dotted black circle centered at the center of mass and with radius given by the maximum distance of the interior discriminant zeros, $i\, Z_0$, which are farthest from the center of mass. 

We see from Figure \ref{fig:peanut-pade} that the Pad\'e poles (red points) accumulate to the discriminant zeros (black points), $i\, Z_0$ that lie inside the planetary boundary. The singularities on the real axis lie outside the planetary boundary and are not related to the Pad\'e poles. As the boundary curvature increases we see that the internal singularities migrate to local maxima of the boundary curvature, while the exterior singularities migrate towards local minima of the boundary curvature. Furthermore, we see that as the curvature increases the region of divergence (a sphere whose cross-section is the dashed black circle) tends to the Brillouin sphere, which touches the outer points of the planetary surface.

\subsection{Roughened planet}
\label{sec:rough}

\begin{figure}[h!]
\centerline{\includegraphics[scale=0.275]{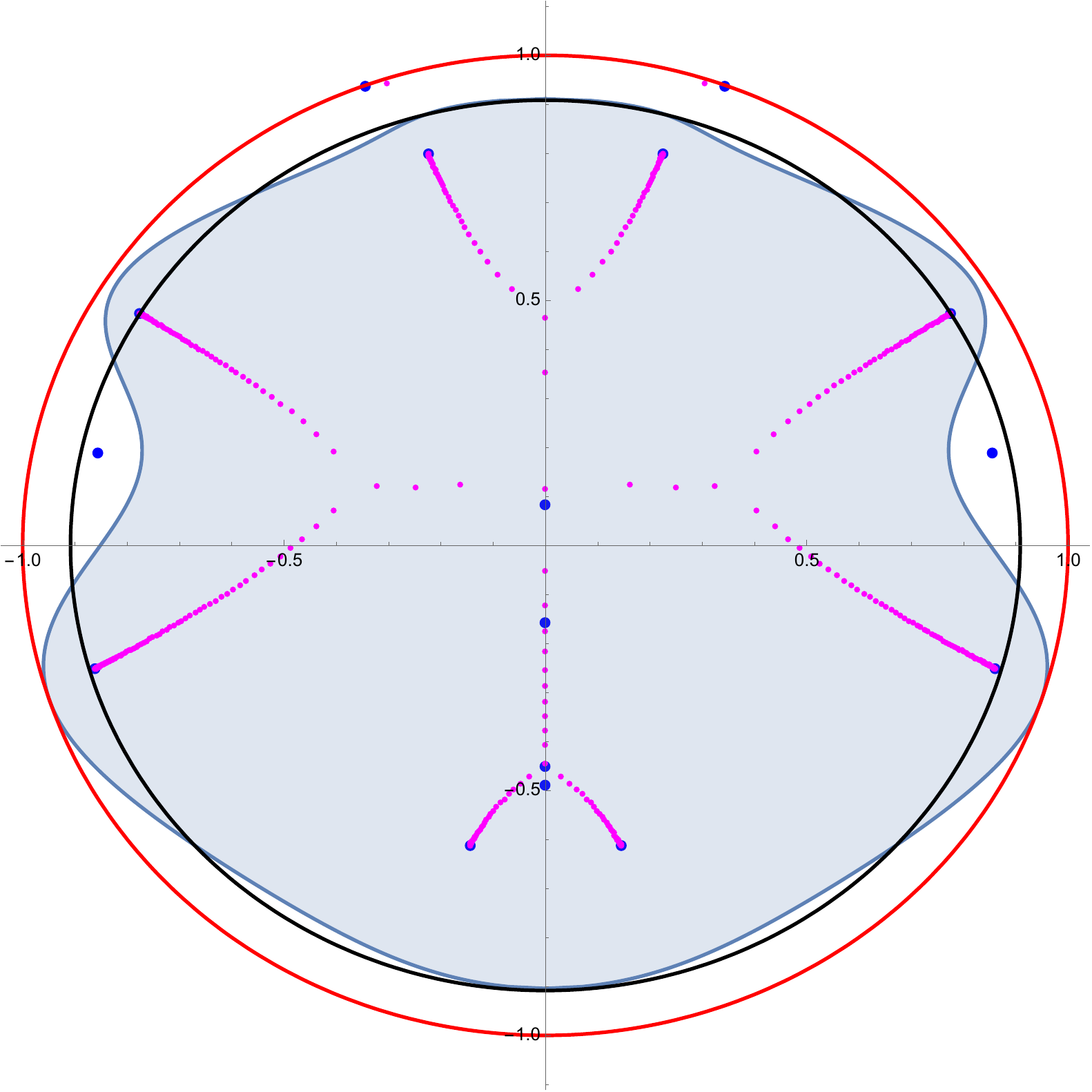} \quad \includegraphics[scale=0.46]{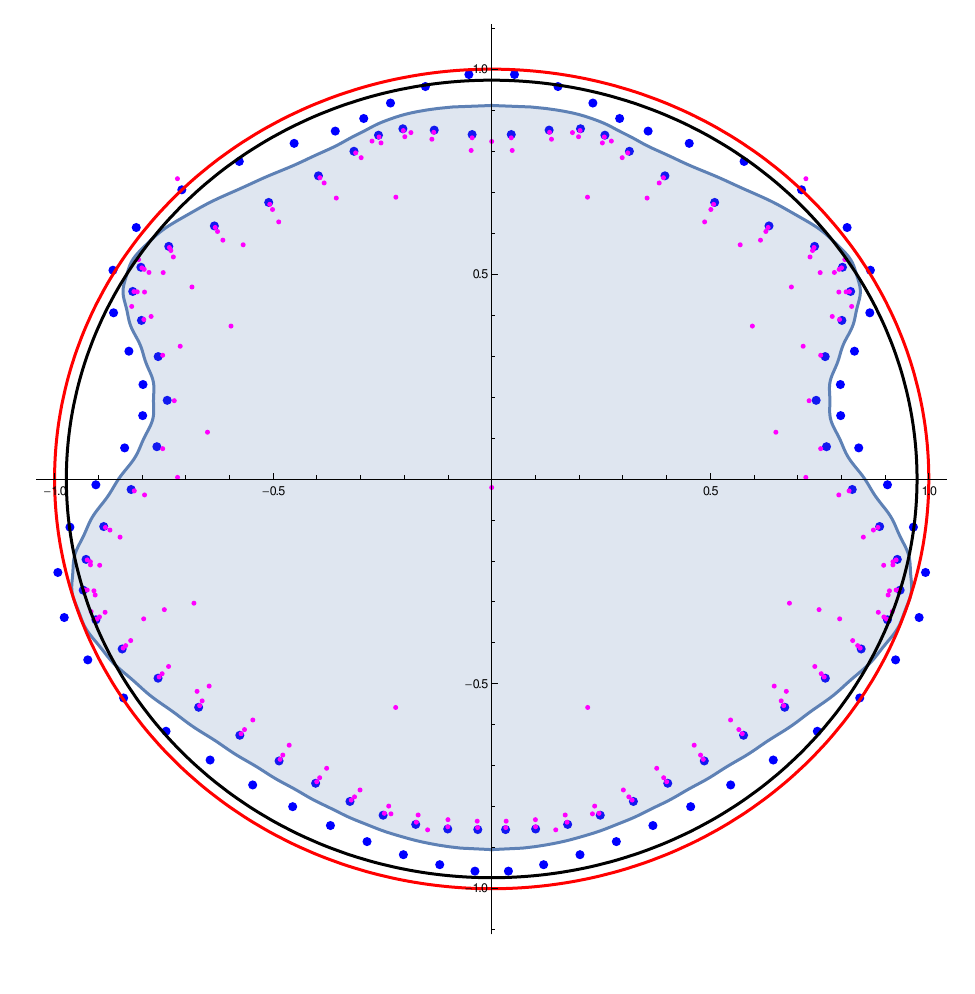} }
\caption{Plots of an order 10 Chebyshev axisymmetric planetary cross-section (blue curves bounding the shaded grey region). The plot on the right has additional small corrugations added to the boundary surface curve. The zeros of the corresponding discriminant are shown as blue dots. The Pad\'e poles are shown as pink dots, accumulating to the discriminant zeros inside the planet. The circle of the Brillouin sphere is shown in red, and the radius of convergence of the SHE is shown as a black circle.}
\label{fig:cheb-pade} 
\end{figure}

To illustrate further the effect of boundary topography on the Pad\'e poles and the associated SHE convergence properties, we consider an axisymmetric planet described by an analytic $s(z)^2$ function that is a high order polynomial, producing smooth peaks and valleys of the boundary curvature. See for example the plot on the left of figure \ref{fig:cheb-pade}, where the gray shaded planetary cross-section is bounded by a smooth (blue) curve. The Pad\'e poles, $i \, Z_0$ are shown as pink points that accumulate to the discriminant zeros shown as blue dots which lie inside the planetary cross-section.  The Brillouin sphere is shown as the red circle, and the radius of convergence of the gravitational potential along the axis of symmetry is shown as a black circle.

We then add a small perturbation to the surface topography in the form of a higher-order polynomial, but with small amplitude. This is shown in the right hand plot in figure \ref{fig:cheb-pade}. This produces an additional small rippled surface roughening. The distortion of the boundary topography is barely visible, but the discriminant is now a much higher order polynomial, so there are many more discriminant zeros, denoted by the blue dots. The Pad\'e poles are  shown as pink dots, and we observe that the Pad\'e poles once again accumulate to the discriminant zeros that lie inside the planet's profile. Moreover, the interior discriminant zeros have migrated to be much closer to the boundary curve, and we also observe that they tend to pair up with discriminant zeros that have migrated to be much closer to the boundary curve but from the exterior. The Pad\'e poles are not sensitive to these exterior singularities, which we interpret to be lying on a higher Riemann sheet. The Brillouin sphere is shown as the red circle, and the radius of convergence of the gravitational potential along the axis of symmetry is shown as a black circle.
As the planetary surface becomes more rough, the singularities tend to the planetary surface, both from the interior and exterior, and the sphere of convergence tends to the Brillouin sphere. This provides a simple physical explanation for the result of \cite{theorem} that for a realistic planet, with non-analytic surface topography, the radius of curvature is determined by the Brillouin sphere. It also shows that adding a small roughening to the surface has a dramatic effect on the Pad\'e pole structure. This means that Pad\'e is a sensitive diagnostic tool for geodesic applications. 

\section{Conclusions}

We have shown that Pad\'e approximants provide a simple and effective framework for analyzing spherical harmonic expansions (SHEs) of the gravitational potential. In particular, the poles of near-diagonal Pad\'e approximants yield direct numerical access to the complex singularities that determine the domain of convergence of the SHEs. Across a range of examples, including smooth, non-smooth, and highly structured planetary topographies, we demonstrate that Pad\'e poles accumulate to the singularities on the first Riemann sheet that govern convergence, while remaining insensitive to those lying on higher sheets. This provides a practical and broadly applicable method for identifying the effective radius of convergence directly from the expansion coefficients, without requiring prior geometric information about the planetary topography or density.

In addition, Pad\'e approximants furnish analytic continuations of the gravitational potential beyond the classical convergence boundary, enabling downward continuation significantly below the Brillouin sphere—indeed, all the way to the planetary topography when the input coefficients are known with sufficient accuracy. At the same time, the examples considered here highlight an inherent limitation: the reliability of Pad\'e continuation depends critically on the precision of the input data, with a sharp threshold beyond which noise severely degrades the approximation. These results suggest that Pad\'e methods offer a valuable complementary tool to existing approaches in geodesy, combining conceptual simplicity with strong numerical performance, while also motivating further work on stability and precision requirements for realistic geophysical data.

While the examples considered here are axisymmetric, chosen for their computational simplicity and the occasional availability of explicit expressions, the method is not restricted to this setting and extends to general geometries. Applications to synthetic models of higher geometric complexity, such as polyhedral models with on the order of $5\times 10^5$ faces, closer to realistic planetary topography, will be addressed in future work.

\section{Acknowledgments}
We thank M. Bevis and R. Costin for discussions. The work of Ovidiu Costin was supported in part by the U.S. National Science Foundation, Division of Mathematical Sciences, Award No. NSF DMS-2206241.

\end{document}